\documentclass{PoS}

\usepackage[sumlimits]{amsmath}

\title{Impact of the Fluctuations in the Extragalactic Background Light on the $\gamma$-ray Attenuation of the Quasars}

\ShortTitle{Impact of the Fluctuations in the EBL on the $\gamma$-ray Attenuation of the Quasars}

\author{Ayman M. Kudoda and Andreas Faltenbacher\\
        School of Physics, University of the Witwatersrand, Braamfontein 2050, South Africa\\
        E-mail: \email{ayman.kudoda@students.wits.ac.za}}

\abstract{This work investigates the impact of the extragalactic background light fluctuations on very high energy $\gamma$-ray spectra from distant Quasars.  We calculate the extragalactic background light spectral energy distribution using a model that builds upon those proposed by Razzaque et al. (2009ApJ.697.483R) and Finke et al. (2010ApJ.712.238F). This model implements the fluctuations in the extragalactic background light based on the fluctuations in the star formation rate density, since both phenomena can reasonably be expected to be correlated.  The fluctuations in the star formation rate are derived from the semi-analytical galaxy catalogue of Guo et al. (2013MNRAS.428.1351G). The determination of the mean, the lower and the upper limits for the scatter of the star formation rate density allows us to compute the corresponding limits on the extragalactic background light spectrum and ultimately the impact of these fluctuations on the $\gamma$-ray optical depth. The model predicts variations of up to $10\%$ between upper and lower limits for the $\gamma$-ray opacity in the energy range less than $100$ GeV for nearby sources. At higher energies the impact is smaller, $\lesssim 5\%$, but still significant even for redshifts $\gtrsim 5$.}

\FullConference{3rd Annual Conference on High Energy Astrophysics in Southern Africa  -HEASA2015,\\
		18-20 June 2015\\
		University of Johannesburg, Auckland Park, South Africa}


\begin{document}

\section{Introduction}

\subsection{Extragalactic Background Light}
\label{sec:ebl}
The Extragalactic Background Light (EBL) is the accumulated light released by the global stellar population from the time of decoupling after the Big-Bang until now.  It contains a wealth of information related to the evolution and the structure of the Universe and its astrophysical components which makes it a topic of great interest.  However, measuring the EBL directly can be difficult for various reasons.  For example, separating it from the zodiacal light of our solar system or from the foreground light of our galaxy \cite{Hauser2001, Dwek2012} poses serious challenges \cite{Costamante2013}.  Nonetheless, indirect measurement of the EBL can be achieved by observing the attenuation of $\gamma$-ray spectra from distant Very High Energy (VHE) sources such as Quasars. As $\gamma$-rays travel through the Universe they interact with EBL photons and produce electron-positron ($e^-$, $e^+$) pairs. This process affects the spectral index of the Quasars in the VHE spectrum regime \cite{Yuan2012}.  

The spectrum of the EBL is distributed between $0.1$ and $1000$ $\mu$m with two distinct humps: a first hump between $0.1$  and $10$ $\mu$m, which is due to direct stellar emission; the contribution from the dust creates a second hump between $10$ and $1000$ $\mu$m. Cosmic Microwave Background (CMB) radiation, X-rays and $\gamma$-rays are not considered to be part of the EBL \cite{Dwek2012} as  they are generated by different production mechanisms.

Since the EBL density in the Universe is generated from galaxies and possibly, primordial stellar populations, the fluctuations in the spatial distribution of these sources lead to fluctuations in the intensity of the EBL \cite{Dwek2012}.  Several studies have been carried  out to investigate these fluctuations.  For instance, Shectman \cite{Shectman1973, Shectman1974} investigated the anisotropy in the optical regime, and Kashlinsky et al. \cite{Kashlinsky1996part1} studied clustering in the near-IR region using Cosmic Background Explorer (COBE) and Diffuse Infrared Background Experiment (DIRBE) Maps,  while the recent study by Furniss et al. \cite{Furniss2015} used the correlation between VHE-emitting sources and cosmic voids along the line of sight to estimate the fluctuations in the attenuation of $\gamma$-rays.  They found an upper limit of $10\%$ for the deviation from the mean for highly attenuated VHE sources.

\subsection{Gamma-ray attenuation by electron-positron pair production}
\label{sec:att}
The interaction of VHE $\gamma$-ray photons with low energy EBL photons producing electron-positron pairs ($\gamma + \gamma' \rightarrow e^+ + e^-$) is the main, and only absorption process we consider here for $\gamma$-ray photons traveling cosmic distances.  This interaction can take place only if the total energy of the two interacting photons is higher than the rest mass of the electron-positron pair.  The cross section of this interaction is described, for instance, in section (II) of \cite{Gould1967}. Based on this we compute the optical depth ($\tau$) for  VHE $\gamma$-ray photons emitted at redshift $z$ and observed at $z = 0$, as given by Eqn. ($17$) in \cite{Razzaque2009}.  

\begin{figure}
\begin{center}
\includegraphics[width=0.65\textwidth]{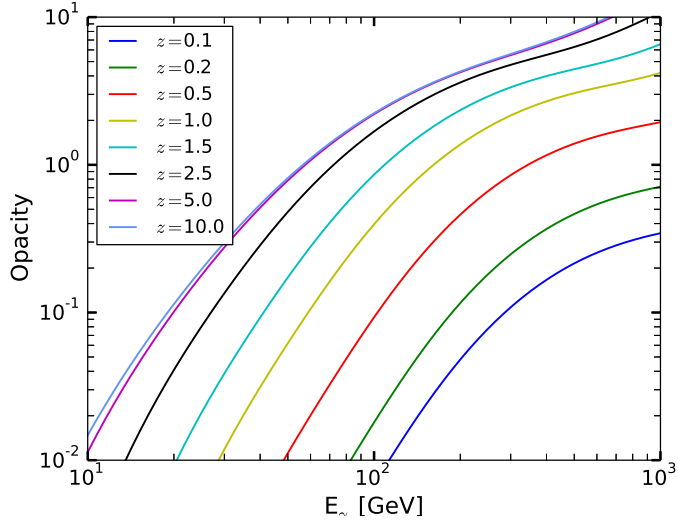}
\caption{Gamma-ray opacity of VHE sources due to interaction with EBL photons. The observed energies, $E_{\gamma}$, range from $0.1$ to $1$ TeV. The opacity is computed for sources at different redshifts ($z=0.1$ to $z=10$) increasing from inner to outer lines.}
\label{opac}
\end{center}
\end{figure}

\section{Modelling the fluctuations in the EBL}
\label{sec:Mod}
In this work we use a semi-analytic approach to consider fluctuations in the EBL. To compute the EBL we employ Eqn. \ref{eq:R} (eq. 12 in \cite{Razzaque2009}) with a \textit{Salpeter B} initial mass function (IMF) for the stellar population. 
\begin{align}\label{eq:R}
  \frac{dN(\epsilon,z=0)}{d\epsilon dV}= &\mathcal{N} \int^\infty_{z=0} dz'' \left|\frac{dt}{dz''}\right| \psi(z'') \int^{M_{\text{max}}}_{M_{\text{min}}} dM \xi(M) \nonumber\\ 
  &\times\int^{z"}_{\text{max}\{0,z_d(M,z')\}}dz' \left|\frac{dt}{dz'}\right| f_{esc}(\epsilon') \frac{dN(\epsilon',M)}{d\epsilon'dt}(1+z'), 
\end{align}
where $\mathcal{N}^{-1}$ is the normalization factor for the IMF $\xi(M)$, $M_{min}$ and $M_{max}$ are $0.1 \rm ~ M_\odot$  and $100 \rm ~ M_\odot$,  $\psi(z)$ is the Star Formation Rate (SFR) at redshift $z$, $z_d$ is the redshift when the star evolves away from the main-sequence, $\frac{dN(\epsilon',M)}{d\epsilon'dt}$ is the total number of emitted photons per time per energy intervals from a star with radius $R$ and temperature $T$, and $f_{esc}(\epsilon')$ is an empirical fitting function for the averaged photon escape fraction from a galaxy, adapted from \cite{Driver2008}. In order to account for the dust re-emission we implement the approach described in Finke et al. \cite{Finke2010}. With this machinery in place we are set to compute the EBL spectrum for $gamma$-ray energies ranging from $0.1$ to $100$ eV.

In order to determine the fluctuations of the EBL we first derive a mathematical description of the fluctuations in the SFR. The rationale behind this approach is that, a region which shows a low SFR for some time will have a lower galaxy population density compared to the cosmic mean. It is self-evident that such regions show lower EBL intensities simply because there is a smaller number of and/or less luminous galaxies in the region. We estimate the fluctuations of the SFR from the semi-analytical galaxy catalogue based on the Millennium simulation Run $7$ (MR$7$) \cite{Guo2013, Lemson2006}, which is sufficiently accurate for this work. The fluctuations are determined by the 5th and 95th percentiles of distribution of SFRs in randomly located spheres of various radii (redshifts). Adding these fluctuations to the EBL model we compute the corresponding limits on the EBL photon density. This allows us to consider two extreme scenarios: 
\begin{itemize}
 \item Firstly, $\gamma$-rays crossing the Universe in over dense regions from the source upto the observer;
 \item Secondly,$\gamma$-rays traveling through under dense regions on their entire path.
\end{itemize}
For this two extreme scenarios we determine the $\gamma$-ray opacity as given by Eqn. ($17$) in \cite{Razzaque2009}.  The full description of our model can be found in \cite[\it{in prep.}]{Kudoda2015}.

\begin{figure}
\begin{center}
\includegraphics[width=0.9\textwidth]{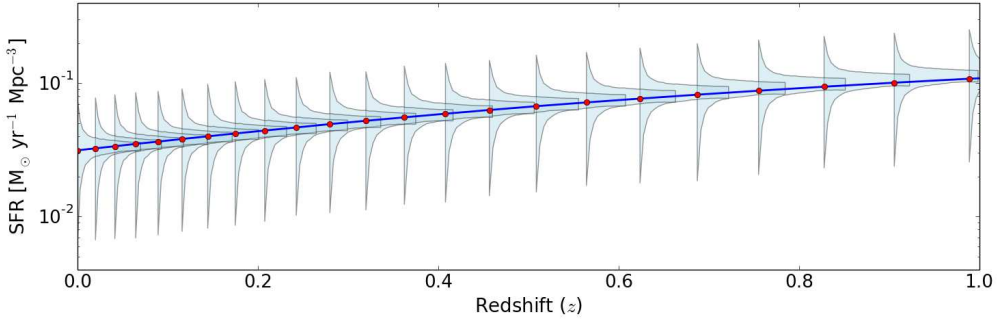}
\caption{The SFR history derived from the semi-analytical galaxy catalogue by Guo et al. (2013) as function of redshift. The red points represent the mean star formation rate in the entire simulation box at the redshifts of the snapshots. The fluctuations are indicated as vertical, ``trumpet-shaped'', shaded regions delineated by the lower 5 and upper 95 percentiles of the distribution of SFRs in randomly located spheres of various radii (redshifts).}
\label{zoom}
\end{center}
\end{figure}

\section{Results}
We calculate the intensity of the $\gamma$-rays (radiation) after they have propagated through the EBL by computing heir interaction probability with the EBL over the redshift range corresponding to the distance of the $\gamma$-ray source.  Figure \ref{opac} shows the optical depth of the $\gamma$-ray interactions with the EBL intensity (see e.g. Eqn. $17$ in \cite{Razzaque2009}).

To estimate the upper and lower limits of the optical depth we use the upper and lower limits of the SFR as shown in Figure \ref{zoom}. This figure demonstrates that the SFR profiles share the same ``trumpet" like shape, i.e. over-plotting the trumpets for different redshifts reveals almost identical shapes for the upper and lower limits. This allows us to use one fitting function for all redshifts. Figure \ref{zoom} also indicates that the value of the SFR density fluctuates strongly, $\pm 50\%$ of the mean density, when the volume is small. For large radii (redshifts) the SFR fluctuations approach the mean. This behaviour is expected as the SFR averaged over large volumes should approach the cosmic mean.

\begin{figure}
\begin{center}
\includegraphics[width=0.7\textwidth]{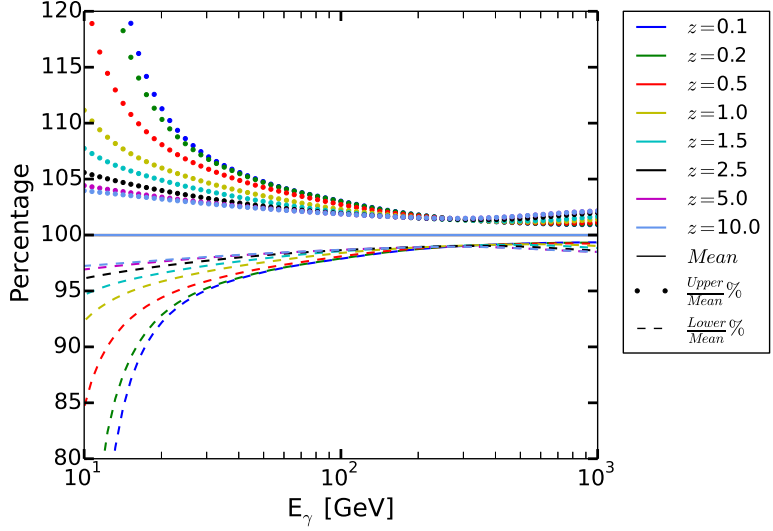}
\caption{The relative difference between the mean $\gamma$-ray opacity and its upper and lower limits as a function of observed $\gamma$-ray energy, $E_{\gamma}$. The doted and the dashed lines are the upper and lower limits. The graphs for all redshifts show a strong increase of the deviation from the mean when the observed energies, $E_{\gamma}$, approach a lower energy limit. This is a consequence of the (mean) optical depth approaching zero for low $\gamma$-ray energies. However, the horizontal part of the graphs indicate a relative difference of $\sim 10\%$ between upper and the lower opacity limits, which will have an impact on the shapes of the high energy parts of $\gamma$ ray depending of whether the rays predominantly encountered over or under-luminous regions.}
\label{Percentage}
\end{center}
\end{figure}

In Figure \ref{Percentage} the impact of the EBL fluctuations on the $\gamma$-ray opacity (optical depth) is shown. The figure displays the relative differences between the opacities based on the extreme cases of 5 and 95 percentiles of the SFR distribution and the mean SFR at different redshifts. This figure indicates that the opacity of VHE $\gamma$-rays varies significantly depending on whether the $\gamma$-rays pass through over or under-dense regions. For example, for $\gamma$-ray sources at lower redshifts, $z\leq 0.5$, the difference between the upper and lower limits is $\sim 10\%$, i.e. the deviation from the mean opacity is $\pm5\%$ at $\gamma$-ray energies $\sim100$ GeV. For $\gamma$-ray sources located at larger redshifts the impact of the fluctuations decreases. For $\gamma$-rays $\geq 100$ GeV the fluctuations change the opacity by $\lesssim 5\%$. The upper and lower limits of the opacity are asymmetric around the mean, i.e. the lower limits are closer to the mean than the upper limits. This is because the fluctuations in SFR tend to be asymmetric in the same sense. The reason for the latter is related to the nature of over densities, they occupy a smaller volume but greatly outnumber under densities in terms of magnitude. These results are also in good agreement with the results derived by Furniss et al. \cite{Furniss2015}.

\section{Conclusion}
Tracking the evolution of the EBL is crucial for the interpretation of $\gamma$-ray observations, but also for deepening the understanding galaxy evolution per se. For that purpose, we used the model introduced by Razzaque et al. \cite{Razzaque2009} and included the dust re-emission discussed in Finke et al. \cite{Finke2010} to determined the evolution of the homogeneous EBL spectrum.

As a new feature, we introduced fluctuations in the EBL model which were assumed to be a result of spatial fluctuations in the SFR. The fitting function for the fluctuations in the SFR density was derived based on MR7 galaxy catalogue \cite{Guo2013}.  Our results show that the fluctuations of the SFR density in small volumes, $\lesssim 25$ Mpc radius, can reach up to $\pm 50\%$ of the mean SFR density. 

The upper and lower limits of the EBL density fluctuations determined at different redshifts were used to compute the impact of the fluctuations on the $\gamma$-ray optical depth. We find variations of up to $10\%$ between the upper and the lower opacity limits for $\gamma$-ray energies less than $100$ GeV from nearby sources. The impact of the EBL fluctuations is smaller ($\lesssim 5\%$) but still significant for VHE $\gamma$-rays from distant sources.

\section*{Acknowledgement}
This research was supported by the National Astrophysics and Space Science Programme (NASSP). The Millennium Simulation (MR$7$) databases  used in this paper and the web application providing online access to them were constructed as part of the activities of the German Astrophysical Virtual Observatory (GAVO).

\end{document}